\documentclass{aa}
\usepackage{epsfig}
\usepackage{psfig}
\usepackage{graphicx}
\usepackage{txfonts}

\usepackage{lscape}
\newcommand{\chisq}{$\chi^{2}$}

\newcommand\G{G67.7+1.8}
\newcommand\CXOU{CXOU195422.97+312902.1}
\newcommand\sten{CXOU195424.75+312824.9}
\newcommand\hard{CXOU195429.82+312834.1}
\titlerunning{X-ray observations of \G }

\begin{document}
\title{Exploring the X-ray emission properties of the supernova remnant \G\ and its central X-ray sources}

\author{C. Y. Hui \and W. Becker}   
\date{Received 12 August 2008 / Accepted 04 December 2008}
\institute{Max-Planck Institut f\"ur Extraterrestrische Physik, 
          Giessenbachstrasse 1, 85741 Garching bei M\"unchen, Germany}

\abstract{ 
We have studied the supernova remnant \G\ with the Chandra X-ray observatory. The 
remnant's X-ray morphology correlates well with the double-arc structure seen 
at radio wavelength. The X-ray spectra of the northern and southern rim of \G\ exhibit 
emission line features of highly ionized metals, which suggests that most of the 
observed X-rays originate in a thermal plasma. We find magnesium , silicon, 
and sulphur overabundant relative to the solar values. Gaussian emission lines 
at $\sim4$ keV and $\sim7$ keV are detected. The $\sim4$ keV line is 
consistent with K-emission lines from $^{44}$Ca and/or $^{44}$Sc whereas the 
$\sim7$ keV line feature may arise from unresolved Fe-K lines. Chandra's sub-arcsecond 
angular resolution allowed us to detect four faint point sources located within $\sim1.5$ 
arc-minutes of the geometrical remnant center. Among these objects, \sten\ and \hard\ 
do not have optical counterparts, leaving them as candidates for a possible compact 
stellar remnant. 

}

\maketitle

\section{Introduction}

Thanks to the progress made in high-energy astrophysics in recent years
other manifestations of neutron stars besides accretion- and rotation-powered 
pulsars have been found. Among the various different categories known today, 
central compact objects (CCOs) are the most enigmatic ones. These sources show 
up as relatively faint X-ray point sources located close to the expansion center of 
their host supernova remnants (SNRs), strongly suggesting that they are indeed the 
compact remnants formed in the supernovae. Despite their apparently
young age, the emission properties of CCOs are found to be very different
from those observed in young rotation-powered pulsars. In the latter the 
X-ray emission is generally dominated by magnetospheric emission whereas 
the X-ray spectra of CCOs can be described well by composite blackbody models 
of temperatures $(T_1,T_2)\sim (3-7)\times 10^{6} K$ and radii $(R_1,R_2)\sim 
0.3-3$~km of the projected blackbody emitting areas that are much smaller than 
the canonical neutron star radius (see Becker 2008 for a review). In contrast 
to the young Crab- and Vela-like pulsars, no plerionic emission has been found 
around CCOs (see Hui \& Becker 2006 and references therein). 

The temporal emission behavior of the known CCOs is another puzzle. So far, 
X-ray pulses have been found from three of the seven known sources. The 
detected periodic signals span a wide range from $P\sim 0.1\,\mbox{s}$ for 
the compact object in SNR  Kes 79 (Halpern et al.~2007) to $P\sim 6.27$ hrs 
in the case of the central point source in RCW 103 (De Luca et al.~2006). 
Different solutions have been proposed to explain their rotational dynamics. 
There is, however, no consensus so far on whether these objects e.g.~were 
born with a slow spin and a weak magnetic field (Halpern et al.~2007) or at 
the other extreme that resembles the magnetars with a very short initial 
spin period and a very strong magnetic field (Li 2007).

Currently, only seven SNRs have been confirmed as hosting a CCO (cf.~Tables 6.3 
and 6.4 in Becker 2008). For a better understanding of their nature,  
the sample of CCOs needs to be increased. We therefore
inspected known supernova remnants for centrally peaked X-ray emission by
reanalyzing the ROSAT all-sky survey (RASS) data. \G\ showed up in RASS data 
to have centrally bright X-ray emission that could potentially be associated 
with a compact source. A faint radio 
source seen in NRAO/VLA Sky Survey (NVSS) data (Condon et al. 1998) 
close to the geometrical center of the remnant 
correlates well with the central X-ray emission (cf.~Fig.1). The latter 
finding raised the question of whether this emission may eventually come 
from a young powerful pulsar and/or its synchrotron nebula.  However, 
searching for a radio pulsar in \G\ did not yield any detection. This 
non-detection placed an upper limit of 800 $\mu$Jy on pulsed radio 
emission at 600 MHz within \G\ (Lorimer et al.~1998). Although this limit 
is not too constraining, it may suggest that the central emission could 
probably be associated with a CCO rather than with a young powerful
rotation-powered pulsar.

\begin{figure*} 
\centerline{\psfig{figure=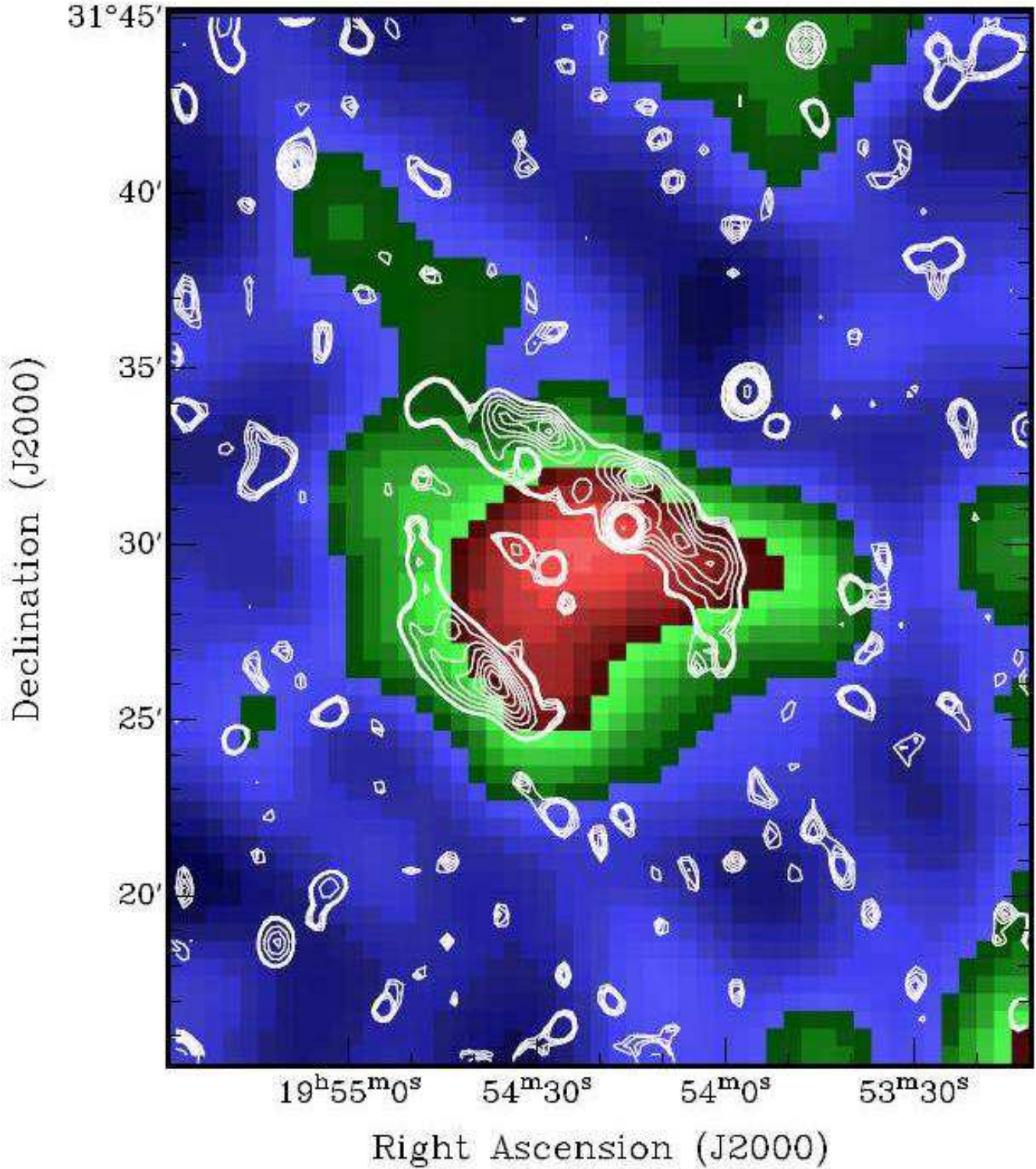,width=18cm,clip=}}
\caption[]{ROSAT PSPC RASS image of G67.7+1.8 ($1.0-2.0$ keV) overlaid 
with contours from the NVSS survey data. The remnant appears center-filled 
in this data and has a bilateral radio arc-like structure. The peak of the 
central X-ray emission is found to be roughly matching the geometric center 
of the remnant. Faint radio sources are seen in the NVSS data near to the 
remnant center.}
\end{figure*} 

\G\ was first discovered in a Galactic plane survey at 327 MHz by Taylor et 
al.~(1992), using the Westerbork Synthesis Radio Telescope. The radio image has 
shown that \G\ has a double arc structure. The angular diameter is $\sim$ $9^{'}$. 
Its distance is quite uncertain. The only distance estimate available comes from
the $\Sigma-$D relation, which puts the remnant at $\sim17$ kpc (Case \& Bhattacharya 
1998). However, optical filaments associated with the northern radio arc of \G\ were found 
by Mavromatakis et al.~(2001) and indicated that the remnants distance might
actually be smaller. In the same work Mavromatakis et al.~(2001) concludes that 
the energy released during the supernova explosion was significantly less 
than the canonical value of $10^{51}$ ergs. 

X-ray emission in the direction towards \G\ was first observed in the ROSAT 
all-sky survey (Mavromatakis et al.~2001). Fitting the data with a thermal 
bremsstrahlung model indicated a plasma temperature in the range of 
$\sim0.2-0.3$ keV.  

In this paper we present a detailed X-ray study of \G\ and investigate the
nature of its centrally peaked emission using Chandra ACIS-I observations. 
Its superior angular resolution enables us to perform a spectro-imaging 
study of the supernova remnant and to search for compact sources and/or 
a pulsar-wind nebula in its central part. 

\section{Observations and data analysis}

\G\ was observed with Chandra in 2007 March $4-5$ with the Advanced CCD Imaging 
Spectrometer (ACIS) using a frame time of 3.2~s. We utilized the whole ACIS-I 
CCD array to obtain an image of the remnant. The data were reprocessed on 2007 
August 7 in order to correct for a systematic aspect offset of 
$\sim0.4"$, possibly caused by a changing thermal environment\footnote{See 
http://cxc.harvard.edu/cal/ASPECT/celmon/ for details}. The total observing 
time on source was $\sim29$ ks. Examining the data for times of high
background we noticed that the observation was affected by soft-proton flares. 
Cleaning the data by removing these flares from the data led to an effective
exposure time of $\sim21$ ks. The analysis was performed in the $0.5-8$ keV
energy band. 

\subsection{Spatial analysis}

A false color image of the ACIS-I data of \G\ is displayed in Figure~2. Comparing 
the X-ray image with the radio contours obtained from the 1.4 GHz 
NVSS data (Condon et al.~1998) shows that the X-ray emission is entirely 
enclosed in the radio arc-like structure (see Figure~3). The center-filled 
X-ray emission and the radio shell-like morphology suggest \G\ belongs to the 
category of ``mixed-morphology" SNRs (Rho \& Petre 1998).  

We have searched for the point sources in the whole ACIS-I data by means 
of the wavelet source detection algorithm and found 36 sources. These sources
are marked by circles in Figure~2. The limiting count rate of the search was 
$2\times10^{-4}$~cts/s. The source positions, positional errors, 
signal-to-noise ratios as well as the estimates of extent 
of all these 36 sources are given in Table~1. Cross-correlating these 
sources with the SIMBAD and NED databases did not result in any identification 
within a search radius of 5 arcsec around each source. 

Twelve of these newly detected sources are located within the supernova remnant. 
Four point-like sources, labeled in Figure~2 with the numbers \#9, \#10, \#11
and \#35 are located within $\sim1.5$ arcmin of the geometrical remnant center. 
Source~\#11 is the brightest among them. Hereafter, we designate 
this source as \CXOU. It is also interesting to 
notice a trail-like radio feature close to the remnant center which has one 
end apparently coincide with source~\#10 (cf.~Figure~3). We designate it as \sten. 
This source appears to be slightly extended, which is also suggested by the 
ratio between its extent and the estimated PSF size (cf. Table~1).  
However, the low significance level of the detection and the patchy envirnoment of 
the remnant emission do not allow one to determine if \sten\ is intrinsically extended. 
In the false colour image, source~\#35 appears to have the hardest X-ray emission 
among these four central point sources (see Figure~2). We designate this source as 
\hard\ hereafter.

 \begin{figure*}
 \centerline{\psfig{figure=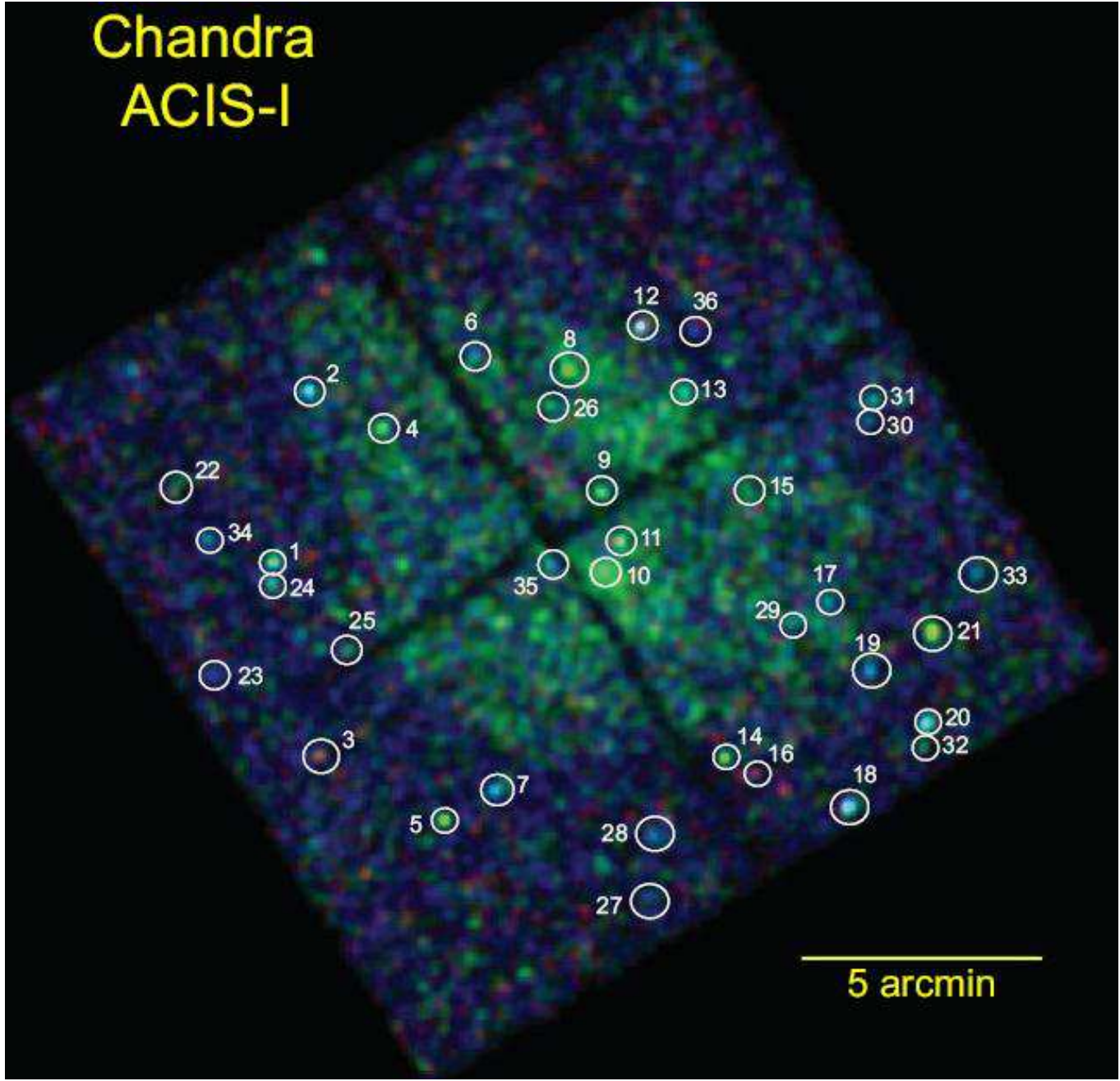,width=18cm,clip=}}
 \caption{Chandra ACIS-I false color image of \G\ (red: 0.3-0.75 keV, green: 0.75-2 keV, 
 blue: 2-8 keV). The binning factor of this image is $2"$. It was adaptively smoothed
 by a Gaussian kernel of $\sigma< 6"$. 36 point sources are detected in the 
 field-of-view (cf.~Table~1).  Source \#11 is designated as \CXOU, source \#10 is \sten. 
 Top is north and  left is east. }
 \end{figure*}

 \begin{figure*}
 \hspace{0.7cm}{\psfig{figure=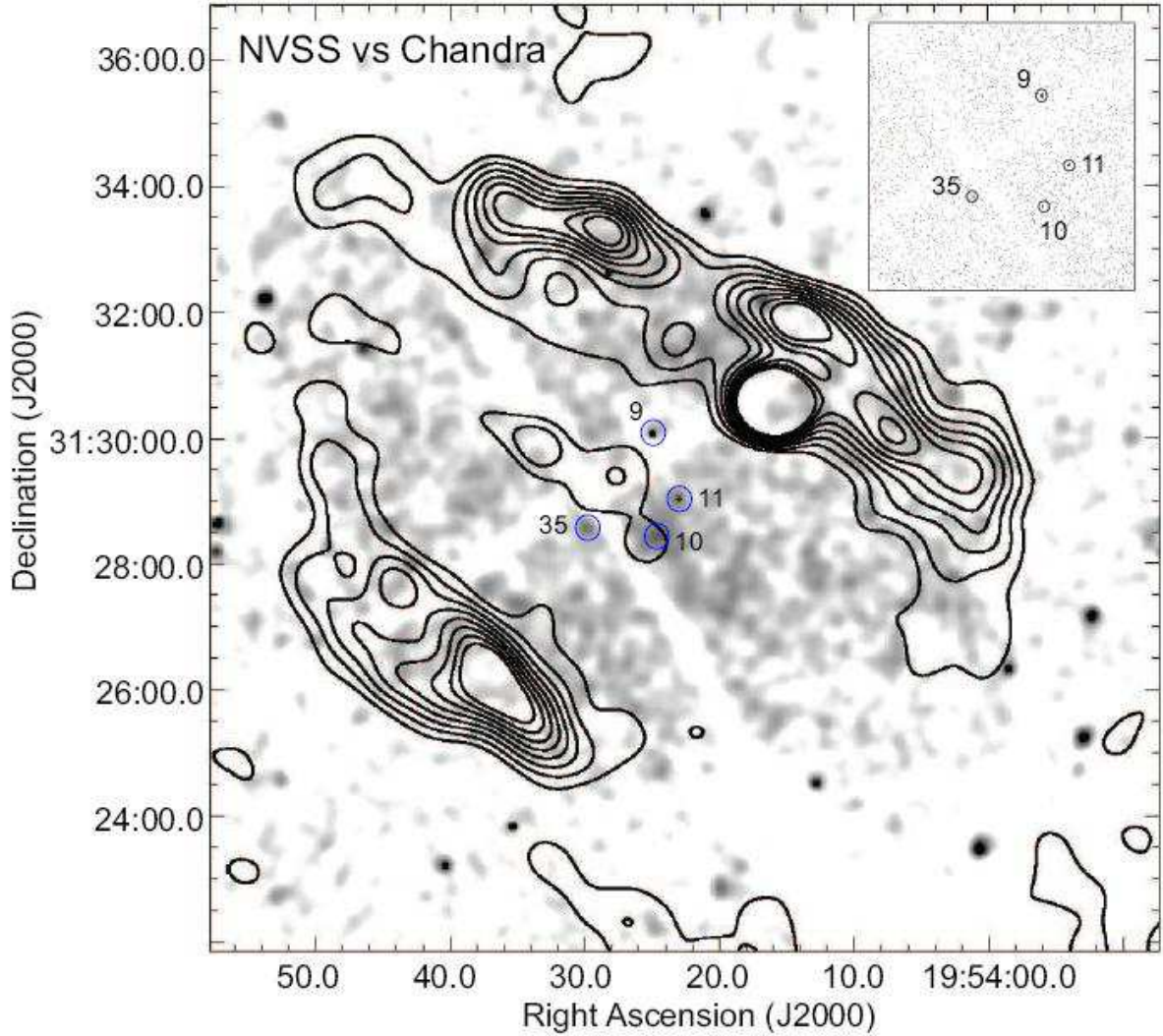,width=16cm,clip=}}
 \caption{$15\times 15$ arcmin field of \G\ as observed by Chandra ACIS-I in 
  $0.5-8$ keV. The radio contours at the levels of $0.6-6$ mJy/beam are from 
  the NVSS data. The blue circles indicate the locations of four point sources 
  near the center of \G. The inset displays the raw image of $4\times 4$ arcmin 
  of the remnant center with the X-ray sources labeled accordingly.}
 \end{figure*}

\begin{table}
\caption{Positions of the X-ray sources in the field of ACIS-I as labeled in Figure~2.  
The corresponding $1\sigma$ positional errors, signal-to-noise ratios as well as 
the estimates of source extent are also tabulated.} 
\resizebox{!}{6cm}{
 \begin{tabular}{l c c c c c c }
  \hline\hline
  Source   &     RA (J2000) &  Dec (J2000)  & $\delta$~RA & $\delta$~dec & S/N$^{~a}$ & PSFRATIO$^{~b}$ \\\hline
           &       h:m:s    &   d:m:s       & arcsec      & arcsec & $\sigma_{G}$ & \\ \hline
1          & 19:54:57.172  & +31:28:36.59   & 0.30    & 0.22 & 13.02 & 1.08  \\
2	   & 19:54:53.728  & +31:32:12.81  &  0.43    & 0.20 & 14.26 & 1.47  \\
3	   & 19:54:52.507  & +31:24:33.74  &  0.70    & 0.28 & 4.71 & 0.80 \\
4	   & 19:54:46.466  & +31:31:26.13  &  0.36    & 0.47 & 6.88 & 2.38  \\
5	   & 19:54:40.352  & +31:23:11.84  &  0.25    & 0.26 & 10.26 & 0.75 \\
6	   & 19:54:37.457  & +31:32:56.14  &  0.54    & 0.33 & 4.71 & 2.19 \\
7	   & 19:54:35.356  & +31:23:49.09  &  0.29    & 0.19 & 9.74 & 0.83 \\
8	   & 19:54:28.210  & +31:32:37.34  &  0.31    & 0.32 & 5.82 & 2.76 \\ 
9	   & 19:54:24.920  & +31:30:04.42  &  0.24    & 0.16 & 10.87 & 5.02 \\
10	   & 19:54:24.745  & +31:28:24.96  &  0.67    & 0.40 & 3.56 & 9.29 \\
11	   & 19:54:22.965  & +31:29:02.08  &  0.14    & 0.17 & 21.18 & 6.13 \\
12	   & 19:54:21.016  & +31:33:33.80  &  0.13    & 0.12 & 24.81 & 1.35 \\
13	   & 19:54:16.717  & +31:32:11.63  &  0.33    & 0.28 & 6.12 & 2.15 \\
14	   & 19:54:12.784  & +31:24:30.48  &  0.37    & 0.52 & 8.36 & 1.78 \\
15	   & 19:54:10.148  & +31:30:01.07  &  0.51    & 0.37 & 2.67 & 1.99  \\
16	   & 19:54:09.741  & +31:24:11.62  &  0.20    & 0.39 & 3.45 & 0.47 \\
17	   & 19:54:02.592  & +31:27:46.74  &  0.27    & 0.34 & 4.59 & 1.03 \\
18	   & 19:54:00.649  & +31:23:28.01  &  0.26    & 0.34 & 16.53 & 0.95  \\
19	   & 19:53:58.504  & +31:26:19.25  &  0.35    & 0.50 & 9.68 & 1.31 \\
20	   & 19:53:52.939  & +31:25:13.26  &  0.28    & 0.40 & 15.18 & 0.90 \\
21	   & 19:53:52.315  & +31:27:08.97  &  0.34    & 0.40 & 13.01 & 1.22 \\
22	   & 19:55:07.125  & +31:30:03.42  &  0.85    & 0.70 & 2.89 & 0.81 \\
23	   & 19:55:03.136  & +31:26:12.22  &  0.82    & 0.51 & 3.28 & 0.96 \\
24	   & 19:54:57.325  & +31:28:09.59  &  0.55    & 0.37 & 6.97 & 1.36 \\
25	   & 19:54:49.830  & +31:26:45.24  &  0.61    & 0.38 & 3.19 & 1.20 \\
26	   & 19:54:30.035  & +31:31:52.60  &  0.73    & 0.44 & 4.22 & 5.44 \\
27	   & 19:54:20.292  & +31:21:33.27  &  0.53    & 0.54 & 4.96 & 0.85 \\
28	   & 19:54:19.929  & +31:22:50.35  &  0.66    & 0.36 & 4.88 & 1.28 \\
29	   & 19:54:06.334  & +31:27:18.46  &  0.42    & 0.43 & 4.03 & 1.68 \\
30	   & 19:53:58.555  & +31:31:33.96  &  0.45    & 0.60 & 4.12 & 1.19 \\
31	   & 19:53:58.106  & +31:32:01.81  &  0.53    & 0.40 & 4.25 & 1.09 \\
32	   & 19:53:53.202  & +31:24:43.93  &  0.60    & 0.57 & 4.45 & 0.68 \\
33	   & 19:53:47.918  & +31:28:22.49  &  0.81    & 0.60 & 3.66 & 1.12 \\
34	   & 19:55:03.855  & +31:29:04.65  &  0.54    & 0.63 & 4.10 & 0.90 \\
35	   & 19:54:29.821  & +31:28:34.14  &  0.67    & 0.49 & 3.78 & 10.81 \\
36	   & 19:54:15.732  & +31:33:27.84  &  0.50    & 0.42 & 4.38 & 1.56 \\
 \hline\\
 \end{tabular}
}
$^{a}$ Estimates of source significance in units of Gehrels error: 
$\sigma_{G}=1+\sqrt{C_{B}+0.75}$ where 
$C_{B}$ is the background counts.\\
$^{b}$ The ratios between the source extents and the estimates of the PSF sizes.  
 \end{table}

\subsection{Spectral analysis of \G}

Before we have extracted the remnant spectra, all point-like sources 
were removed. The spectra were then extracted from the elliptical-shaped 
regions illustrated in Figure~4. These regions cover
both the northern and the southern radio shells. The background 
spectra for the corresponding CCD chips were extracted from the 
boxed regions marked in Figure~4. We utilized 
the tool SPECEXTRACT in the data reduction package CIAO~3.4.1 with the 
calibration data CALDB~3.4.1 to extract the spectra and to compute the 
response files. After background subtraction there are 1028 and 610 net 
counts avaliable for the spectral analysis of the northern rim 
(i.e.the sum of two northern elliptical regions) 
and southern rim (i.e. the sum of two southern elliptical regions), respectively. 
The background contributions are found to be 
$\sim60\%$ in both, the northern and southern rims. We binned all the 
remnant spectra dynamically so as to have at least 50 counts per bin. All the 
spectral fittings 
were performed with XSPEC 11.3.2. The parameters of the best-fit 
models are summarized in Table~2. All quoted errors are $1\sigma$ 
for 1 parameter of interest.

 \begin{figure}
 \centerline{\psfig{figure=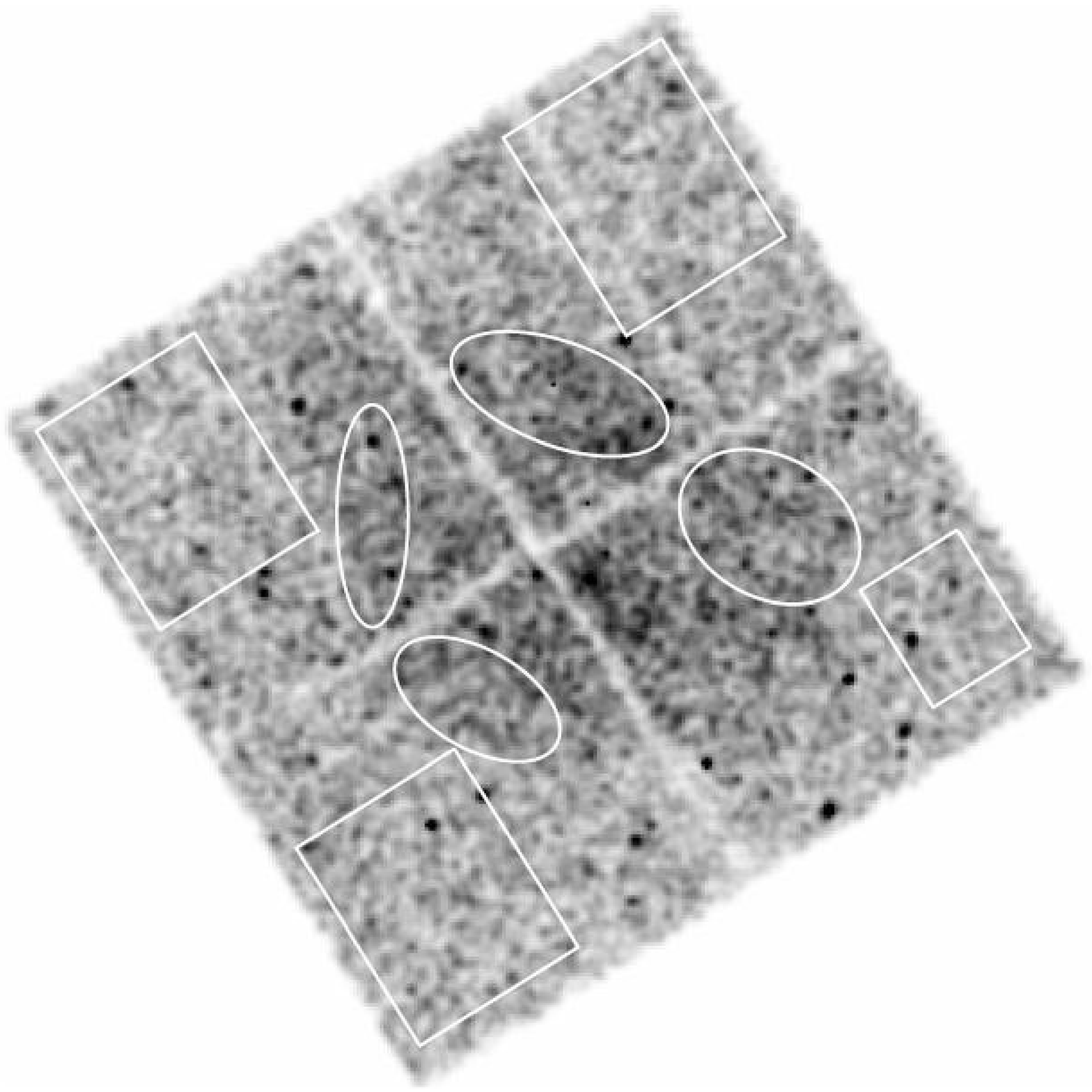,width=9cm,clip=}}
 \caption{Illustration of the elliptical-shaped regions used to extract 
 the spectra from the northern and southern rim of \G. The background
 spectra were extracted from the rectangular regions.}
 \end{figure}

\subsubsection{The northern rim}

The X-ray spectrum of the northern rim is shown in Figure~5 which appears to be 
rich with line features. We attempted to fit the data with an absorbed 
collisional ionization equilibrium (CIE) plasma model (XSPEC model: VEQUIL). 
To examine the abundance of metals, we thawed the parameters individually 
to see whether the goodness-of-fit can be improved and/or the abundance 
significantly deviates from the solar value. However, this single component 
model cannot depict the data beyond 3 keV (with \chisq=45.32 for 33 D.O.F.) 
and requires an unreasonably overabundance of iron and calcium (i.e. hundred 
times of the solar values). 

We proceeded to examine whether the excess in the residual can be modeled by a
second thermal component. Adding another CIE component with the solar abundances, 
we found no improvement in the goodness-of-fit (\chisq=47.71 for 31 D.O.F). 

After examining the residual carefully, we found that combining a CIE model 
with two additional Gaussian components can model the observed spectrum. The 
composite model can describe the data within $0.5-8$ keV very well: 
\chisq=25.41 for 29 D.O.F. (cf.~the lower panel of Figure~5). All the best-fit 
parameters are tabulated in Table~2. 

The best-fit model yields a hydrogen column density of 
$n_{H}=(4.1\pm0.9) \times10^{21}$ cm$^{-2}$. Based on the H$\alpha$/H$\beta$ line ratio, 
Mavromatakis et al. (2001) inferred an optical extinction of $\sim2$. This value implies 
a hydrogen column density of $\sim4\times10^{21}$~cm$^{-2}$ towards \G\ (Predehl \& Schmitt 
1995), which is in good agreement with our best-fit value. For comparison, the total galactic 
neutral hydrogen column density towards \G\ is $\sim10^{22}$ cm$^{-2}$ 
(Kalberla et al.~2005; Dickey \& Lockman 1990). The plasma temperature is found to 
be $T=6.6^{+0.5}_{-0.7}\times10^{6}~K$. For the metal abundance, 
our analysis suggests that magnesium, silicon and sulphur are overabundant with respect 
to the solar values (Mg:2.6$^{+0.9}_{-0.7}$, Si:2.8$^{+1.5}_{-1.2}$, 
S:13.6$^{+8.7}_{-6.9}$). The best-fit parameters imply the unabsorbed flux 
to be $f_{x}=6.7\times10^{-13}$ ergs cm$^{-2}$ s$^{-1}$ in $0.5-8$ keV. 

For the two additional Gaussian line features, the centroids of the line energy were found to 
locate at $E_{1}=4.0\pm 0.2$ keV and $E_{2}=7.3^{+3.2}_{-0.5}$ keV. The FWHMs of the 
line profiles are $\sigma_{1}=0.3^{+0.2}_{-0.1}$ keV and $\sigma_{2}=0.9^{+2.0}_{-0.3}$ keV 
respectively. The best-fit line fluxes of the features are $f_{\rm line_{1}}=3.3^{+1.4}_{-1.3}
\times10^{-6}$ photons cm$^{-2}$ s$^{-1}$ and $f_{\rm line_{2}}=2.9^{+12.7}_{-1.4}\times10^{-5}$ 
photons cm$^{-2}$ s$^{-1}$ respectively. The possible physical nature and the significance 
of the emission line features are discussed in \S3. 


We have further examined whether a non-equilibrium ionization collisional plasma model 
(XSPEC model: VNEI) can provide a better fit than a CIE model. With a composite model 
of VNEI model plus two Gaussian components, We do not found any improvement 
in the goodness-of-fit (\chisq=25.41 for 28 D.O.F). All the best-fit parameters are consistent 
with those inferred from the CIE model. Moreover, the inferred ionization timescale is 
$\sim4.5\times10^{13}$~s~cm$^{-3}$ which implies the condition of CIE (Borkowski et al. 2001). 
This justifies the validity of adopting the CIE model in our analysis.

The detection of the emission line features prompted us to investigate their spatial distribution 
in \G\ by producing an X-ray image in the energy band of $3-8$ keV. However, owing to the high 
background contribution and the relatively limited photon statistics, we cannot identify any 
significantly excess emission from the background in the image. 

 \begin{figure}
 \centerline{\psfig{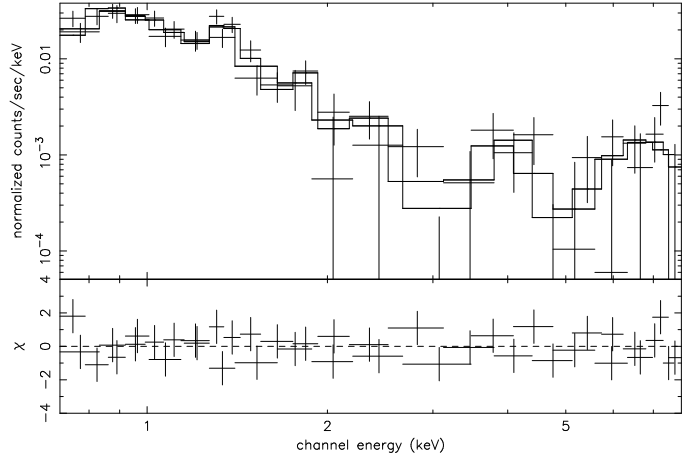}}
 \caption{X-ray spectrum of the emission from the northern rim
 ({\it upper panel}) observed by Chandra's ACIS-I2 and I3 CCDs. The
 residuals are from fitting an absorbed collisional ionization equilibrium 
 plasma model plus two additional Gaussian components ({\it lower panel}). }
 \end{figure}

\subsubsection{The southern rim}

The X-ray spectrum of the southern rim is shown in Figure~6. The spectrum
appeared to be similar to what we observed from the northern rim, so that we fitted
the data again with a CIE model. We have also examined the metal abundances 
as described in the previous subsection. However, the single component CIE model 
is not able to describe the data beyond $\sim4$ keV (with \chisq=38.82 for 23 D.O.F.). 
It also requires an unreasonably high abundance of iron. 

Inspecting the fit residuals reveals 
the existence of a Gaussian component centered at $\sim7$ keV. 
However, modeling the data with the line energy of the Gaussian profile as 
a free parameter did not yield a stable fit. In particular, the centroid 
of the line cannot be constrained. This can be ascribed to the relatively 
low photon statistics of the southern rim spectrum. We therefore proceeded 
to fit the spectrum with the centroid of the line fixed at 7 keV. We found 
that the CIE model complemented with a Gaussian line at 7 keV can describe 
the data fairly well: \chisq=31.37 for 23 D.O.F.. Different from the case of 
the northern rim, we did not find any excess around $\sim4$ keV in the residual. 

The best-fit model yields a hydrogen column density of 
$n_{H}=5.9^{+1.2}_{-1.1}\times10^{21}$ cm$^{-2}$ and a plasma temperature 
of $T=(7.0\pm0.7)\times10^{6}~K$. Within the $1\sigma$ errors 
these values are consistent with those inferred from the spectral fit of the 
northern rim. The overabundance of magnesium, silicon and sulphur with respect 
to the solar values are Mg:$1.7^{+1.1}_{-0.7}$, Si:$4.7^{+2.6}_{-1.7}$, 
S:$16.2^{+11.7}_{-7.4}$. The best-fit parameters imply the unabsorbed flux 
to be $f_{\rm plasma}=4.2\times10^{-13}$ ergs cm$^{-2}$ s$^{-1}$ in $0.5-8$ keV.

For the Gaussian line feature at 7 keV, the FWHM of the profile is found 
to be $1.1^{+0.4}_{-0.3}$ keV. The line flux of this  feature is 
$f_{\rm line}=(1.4\pm0.5)\times10^{-5}$ photons cm$^{-2}$ s$^{-1}$. 


Adopting the VNEI model does not provide a better description of the data than the CIE model. 
The composite model of VNEI component plus the Gaussian fixed at 7 keV results in a 
goodness-of-fit of \chisq=31.36 for 22 D.O.F.. All the best-fit parameters are consistent 
with those inferred from the CIE model. The ionization 
timescale is found to be $\sim4.5\times10^{13}$~s~cm$^{-3}$ which is similar to that inferred 
from the northern rim. Therefore, the CIE condition is justified and we will not further consider 
the VNEI model.

 \begin{figure}
 \centerline{\psfig{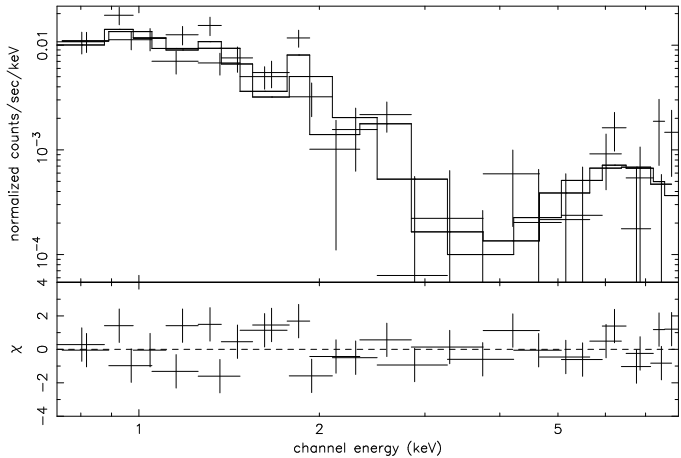}}
 \caption{X-ray spectrum of the emission from the southern rim 
 ({\it upper panel}) observed by Chandra's ACIS-I0 and I1 CCDs. The 
 residuals are from fitting an absorbed collisional ionization equilibrium 
 plasma model plus an additional Gaussian component 
 with the centroid fixed at 7 keV ({\it lower panel}). }
 \end{figure}

\subsection{Spectral analysis of the central point sources}

\CXOU\ (source \#11) is the brightest one among the four point sources 
near to the remnant center. We extracted its spectrum from a circular region 
with a radius of 10 arcsec centered at the source position (cf. Table~1). 
The background spectrum was extracted from a nearby region from a circle of
10 arcsec radius centered at RA=$19^{\rm h}54^{\rm m}23.234^{\rm s}$,
Dec=$31^{\circ}29'22.75"$ (J2000). After background subtraction, 54 net
counts were avaliable for the spectral analysis of the source.
Since the location of \CXOU\ is close to the edge of
ACIS-I3 CCD, we computed the response files manually with the CIAO tools MKARF and MKRMF.
The spectrum was binned dynamically so as to have at least 5 counts per bin. 
In view of the small photon statistic of these sources, we adopted the $C-$statistic 
(Cash 1979) for all the fittings. To better constrain the parameters, we fixed 
the hydrogen column density at the optical extinction inferred value 
(i.e. $4\times10^{21}$~cm$^{-2}$). 

Fitting with a power-law model results in a reasonable fit that yields a 
photon index of $\Gamma=2.7\pm0.4$ and a normalization at 1 keV 
of $1.4^{+0.4}_{-0.3}\times10^{-5}$ photons keV$^{-1}$ cm$^{-2}$ s$^{-1}$. 
The best-fit column density is consistent with the values inferred from the spectra 
of the supernova remnant emission within the $1\sigma$ error bound. The unabsorbed flux 
deduced for the best-fit power-law model parameters is $f_{x}=4.5\times10^{-14}$ 
ergs cm$^{-2}$ s$^{-1}$ in $0.5-8$ keV. 

A blackbody model can describe the spectrum of this source equally well. 
The best-fit model implies a 
temperature of $T=4.6^{+0.7}_{-0.6}\times10^{6}~K$. The radius of the
projected blackbody emitting area is in the range $\sim170-310$ m and 
$\sim410-750$ m for the adopted distances of 7 kpc and 17 kpc, respectively 
(see \S3 for the discussion of the remnant distance). The unabsorbed flux 
deduced for the best-fit blackbody model is $f_{x}=2.6\times10^{-14}$ 
ergs cm$^{-2}$ s$^{-1}$ in $0.5-8$ keV. 

For the other three fainter central X-ray sources  
it is interesting to compare their brightness and hardness 
with those of \CXOU. To do so, we have prepared their spectra 
and the response files in the same way as we did for \CXOU. Fixing 
the column density at $n_{H}=4\times10^{21}$ cm$^{-2}$ we obtained 
the photon indices by fitting a power-law model to their spectra. 
The fitted parameters are summarised in Table~3. The photon 
index provides a measure of the hardness of these X-ray sources. Whereas 
sources \#9 and \sten\ (source~\#10) are as soft as \CXOU, \hard\ (source \#35) appears 
to show harder X-ray emission. We have also computed the 
absorption-corrected fluxes from the inferred power-law parameters 
which are given in Table~3.

\subsection{Optical identifications of the central point sources}

Given the limited photon statistics of the central point sources, the spectral 
analysis is not very constraining. Even for the brightest object \CXOU, we cannot 
distinguish its emission nature between the thermal and the non-thermal scenarios. 
To investigate if these sources are promising neutron star candidates we proceeded to 
search for their optical counterparts by utilizing the USNO-B1.0 catalogue (Monet et al. 2003). 

For the search regions, we combine the positional errors of each source with the pointing 
uncertainty of the spacecraft. The uncertainty can be estimated
from the distribution of aspect offset for a sample of point sources with accurately
known celestial positions\footnote{http://cxc.harvard.edu/cal/ASPECT/celmon/}. 
There is $68\%$ of 70 sources imaged on ACIS-I have offsets smaller than $\sim0.4$~arcsec. 
We adopted this value as the astrometric uncertainty and added to the quoted positional errors (cf. Table~1) 
in quadrature for each coordinate. 

For source~\#9, we have identified an optical counterpart at 0.39 arcsec away from its X-ray position. 
It has a magnitude of $B=15.93$ and $R=14.69$ which implies an X-ray-to-optical flux ratio to be  
$f_{X}/f_{\rm opt}\sim10^{-3}$. This ratio suggests source~\#9 is likely to be a field star which typically has 
a ratio $f_{X}/f_{\rm opt}<0.3$ (Maccacaro et al. 1988). 

For \CXOU, an optical counterpart with $B=15.92$ and $R=14.43$ is consistent
with its X-ray position. For this source the X-ray-to-optical flux ratio is at 
the level of $\sim10^{-3}$ which is in agreement with what is expected for a
field star. 

We do not find any optical counterpart for \sten\ and \hard. Adopting the limiting magnitude in the 
USNO-B1.0 catalogue to be 21 (cf. Monet et al. 2003), the upper limit of $f_{X}/f_{\rm opt}$ is found to be 
$>0.1$. This upper limit is not particular constraining. A deep optical observation 
is required to differentiate the nature of these two sources from the star with $f_{X}/f_{\rm opt}<0.3$ and 
active galactic nuclei with $f_{X}/f_{\rm opt}<50$ (Maccacaro et al. 1988; Stocke et al. 1991). 

\begin{table}
\caption{Spectral parameters of the best-fit model inferred from the Chandra observed spectra of \G.}
\begin{center}
 \begin{tabular}{l c c }
  \hline\hline\\[-2ex] 
  & The Northern Rim & The Southern Rim \\\\[-2ex]
\hline\\[-2ex]
 $n_{H}$ ($10^{21}$ cm$^{-2}$)       &  $4.1\pm0.9$               & $5.9^{+1.2}_{-1.1}$\\[1ex]
 $T$ ($10^{6}~K$)                    &  $6.6^{+0.5}_{-0.7}$       & $7.0\pm0.7$ \\[1ex]
 Mg                                  &  $2.6^{+0.9}_{-0.7}$       & $1.7^{+1.1}_{-0.7}$ \\[1ex]
 Si                                  &  $2.8^{+1.5}_{-1.2}$       & $4.7^{+2.6}_{-1.7}$ \\[1ex]
 S                                   &  $13.6^{+8.7}_{-6.9}$      & $16.2^{+11.7}_{-7.4}$ \\[1ex]
 Norm$_{\rm cie}$ ($10^{-4}$)$^{a}$ & $1.4^{+0.6}_{-0.4}$        & $1.0^{+0.5}_{-0.4}$ \\[1ex]
\hline\\[-2ex]
 $E_{1}$ (keV)                        & $4.0\pm0.2$               & - \\[1ex]
 $\sigma_{1}$ (keV)                   & $0.3^{+0.2}_{-0.1}$       & - \\[1ex]
 Norm$_{\rm gauss1}$ ($10^{-6}$)$^{b}$& $3.3^{+1.4}_{-1.3}$       & - \\[1ex]
\hline\\[-2ex]
 $E_{2}$ (keV)                             & $7.3^{+3.2}_{-0.5}$  &  7 (fixed)  \\[1ex]
 $\sigma_{2}$ (keV)                        & $0.9^{+2.0}_{-0.3}$  & $1.1^{+0.4}_{-0.3}$  \\[1ex]
 Norm$_{\rm gauss2}^{b}$ ($10^{-5}$)$^{b}$ & $2.9^{+12.7}_{-1.4}$   & $1.4\pm0.5$  \\[1ex]
 \hline\\[-2ex]
\chisq\                   &  25.41  & 31.37  \\[1ex]
D.O.F.                    &  29  &  23 \\[1ex]
 \hline
 \end{tabular}
 \end{center}
$^{a}$ {\scriptsize The normalization of CIE model is expressed as 
$(10^{-14}/4\pi D^{2})\int N_{e}N_{H}dV$ where $D$ is the source distance 
in cm and $N_{e}$ and $N_{\rm H}$ are the post-shock electron and hydrogen densities in cm$^{-3}$.}\\
$^{b}$ {\scriptsize The normalization of the Gaussian model is in unit of photon cm$^{-2}$ s$^{-1}$.}
 \end{table}


\begin{table}
\caption{Spectral characteristics of the central X-ray sources within the supernova remnant \G.}
\begin{center}
 \begin{tabular}{c c c c}
  \hline\hline
Source    &   net rate        & $\Gamma$ &            $f_{x}$             \\
{}        &  ($10^{-3}$ cts/s)&    {}        &   (ergs cm$^{-2}$ s$^{-1}$)    \\\hline\\[-2ex]
9         &  $1.17\pm0.24$   & $3.2^{+1.0}_{-0.8}$    & $2.8\times10^{-14}$\\[1ex]
10        &  $0.42\pm0.26$   & $2.7^{+2.4}_{-1.7}$    & $9.8\times10^{-15}$\\[1ex]
11        &  $2.54\pm0.39$   & $2.7\pm0.4$            & $4.5\times10^{-14}$\\[1ex]
35        &  $0.75\pm0.23$   & $1.3^{+0.7}_{-0.5}$    & $1.2\times10^{-14}$\\[1ex]
 \hline
 \end{tabular}
 \end{center}
 \end{table}

\section{Summary and Discussion}

We have performed a detailed spectro-imaging X-ray study of the supernova remnant \G\ with 
Chandra. Various properties of \G, including distance, age and explosion energy, are still 
poorly constrained. The type of the supernova and the nature of the progenitor are also 
unknown. With the Chandra observation it became possible for the first time to shed some 
more detailed light on the X-ray emission nature of \G\ and its central point sources.

The spectra of the remnants northern and southern rims can be described
with a single temperature CIE model plus additional Gaussian components. 
Our analysis suggests a plasma temperature of $\sim7\times10^{6}~K$. Assuming that the 
ions and electrons are fully equilibrated in the shock, the temperature 
implies that the blast wave has a velocity of $\sim800$ km/s which is 
at the same order of magnitude as inferred from the RASS data (Mavromatakis 
et al. 2001). Although the single temperature plasma model describes our 
observed spectra very well, most supernova remnants require multi-temperature 
models to depict their X-ray spectra (e.g.~Gaensler et al.~2008). A composite 
model can describe the shocked ejecta and the swept-up material from the 
surrounding with the components of high temperature and low temperature, 
respectively. However, for a more detailed modeling of the remnant 
spectrum, data with improved photon statistics are badly needed. 

Utilizing the $\Sigma-D$ relation, Case \& Bhattacharya (1998) estimated 
the distance to \G\ to be $\sim17$ kpc. However, the large uncertainties 
on the proportionality factor and the exponent of the $\Sigma-D$ relation 
make the distance poorly constrained and results in a range of $7-27$ kpc. 
On the other hand, the observation of the optical filaments suggests a 
distance less than 17 kpc, otherwise the detection of the optical 
radiation would be very difficult to explain in view of the measured 
interstellar extinction (Mavromatakis et al. 2001). 
In the following calculations we assume the distance to be in the range of $7-17$ kpc. 

The emission measure inferred from the spectral fits of the remnant 
emission allows us to estimate the hydrogen density, $N_{\rm H}$, 
in the shocked regions. Assuming the post-shock densities of hydrogen and 
electrons are uniform in the extraction region, the normalization 
of the CIE model can be approximated by $N_{\rm cie}\simeq 
10^{-14}N_{\rm H}N_{e}V/4\pi D^{2}$, where $D$ is the distance
to \G\ in $cm$ and $V$ is the volume of interest in units of $cm^{3}$. 
Assuming a geometry of oblated spheroid, the volume of interest for 
the northern rim and the southern rim are $3.9\times10^{55}
d^{3}_{\rm kpc}$ cm$^{3}$ and $2.6\times10^{55}d^{3}_{\rm kpc}$ cm$^{3}$, 
respectively. Here $d_{\rm kpc}$ is the remnant distance in units of kpc. 
Assuming a fully ionized plasma, $N_{\rm H}N_{e}$ can be approximated by 
$0.7N_{\rm H}^{2}$. The best-fit normalizations imply a post-shock hydrogen 
density of $N_{\rm H} \sim0.06-0.1$ cm$^{-3}$ for the assumed range of distances. 

Assuming \G\ is in a Sedov phase, the shock temperature can be estimated by:
\begin{equation}
T_{s}\simeq 8.1\times 10^{6}\left(\frac{E_{51}}{N_{\rm H_{-1}}}\right)^{\frac{2}{5}}t_{4}^{-\frac{6}{5}}~K
\end{equation}

\noindent where $t_{4}$, $E_{51}$ and $N_{\rm H_{-1}}$ are the time after the explosion in units of $10^{4}~years$,  
the released kinetic energy in units of $10^{51}~ergs$ and the post-shock hydrogen 
density of $0.1~cm^{-3}$ respectively (McCray 1987). 
Mavromatakis et al.~(2001) has argued the explosion energy should be significantly less 
than the canonical value of $10^{51}$ ergs. Adopting an explosion energy of 
$E=10^{50}$ ergs and the plasma temperature inferred from our spectral analysis 
as an estimate of $T_{s}$ (i.e.~$\sim7\times10^{6}~K$), the range of 
$N_{\rm H}\sim0.06-0.1$ cm$^{-3}$ yields the age of \G\ to be $\sim5000-6000$ 
yrs. On the other hand, an age bracket of $\sim11000-13000$ yrs is suggested 
from $E=10^{51}$ ergs. 

Taking the estimates of the age, explosion energy and the density of the 
interstellar medium, we calculate the theoretical size of the remnant 
by (Culhane 1977): 

\begin{equation}
R_{s}\simeq 21.5\left(\frac{E_{51}}{N_{\rm H_{-1}}}\right)^{\frac{1}{5}}t_{4}^{\frac{2}{5}}~{\rm pc}. 
\end{equation}

From the X-ray image, we estimated the angular radius of the remnant to be 
$\theta_{s}\simeq5$ arcmin. Comparing $R_{s}$ and $\theta_{s}$, we can 
estimate the distance by $d=R_{s}/\theta_{s}$.  For an explosion energy 
of $E=10^{50}$ ergs, the distance is estimated to be in a range of 
$\sim7-12$ kpc for $N_{\rm H}=0.06-0.1$ cm$^{-3}$ and $t=5000-13000$ yrs. 
On the other hand, if $E=10^{51}$ ergs is adopted, it implies a distance 
bracket of $\sim12-20$ kpc for the same ranges of $N_{\rm H}$ and $t$ in 
the previous calculation. 

As aforementioned, the optical extinction suggests that the distance is
at the lower side. On the basis of our estimation, for a preferable 
distance much smaller than $17$ kpc, we favor a scenario 
that the explosion is less energetic than a canonical supernova 
(i.e.~$E<10^{51}$ ergs), which is consistent with the conclusion 
drawn by Mavromatakis et al.~(2001) from their optical study. 

Assuming that the remnant \G\ is the result of a low energy supernova 
explosion has interesting implications for the properties of its progenitor.
In the context of the current understanding in stellar evolution, a 
supernova with kinetic energy of $\sim10^{50}$ ergs can be 
a result of two different evolutionary tracks. 

In the first scenario, a Oxygen-Neon-Magnesium (ONeMg) core can be formed 
before Neon and subsequent nuclear burning take place. If the neutrino 
cooling is efficient enough, the core temperature will be reduced and 
prevents further nuclear fusion. When the ONeMg core reaches the Chandrasakar 
mass, the electron degenerate pressure is no longer able to support the 
core. Furthermore, electron capture by $^{24}$Mg and $^{20}$Ne will further 
reduce the electron degenerate pressure in the core and trigger the 
core to collapse (Miyaji et al.~1980; Guti\'errez, Canal \& Garca-Berro 2005). 
This is known as the electron-capture supernova. Most simulations have 
shown that this type of explosion has a rather low energy (see Eldridge, 
Mattila, \& Smartt 2007). The progenitor's mass of an electron-capture 
supernova is limited in a narrow range of $\sim6-8$ M$_{\odot}$ (Eldridge, 
Mattila, \& Smartt 2007). For more massive stars, they will go through 
all stages to silicon burning. In less massive stars, ONeMg cores cannot 
be formed. 

There is another possibility to produce a low energy supernova. The exact
evolution of a collapsing star after neutrino re-heating depends on the
rate of early infall of stellar material on the collapsed core and on the
binding energy of the envelope. If both are large, which is the case in
high-mass stars, the energy required to accelerate and heat up the ejecta
is not available, preventing a successful explosion and resulting in a
supernova which appears to be under-energetic (see Heger et al.~2003; 
Eldridge \& Tout 2004). The progenitor of this evolutionary track is 
likely to be $\gtrsim20M_{\odot}$.

We have also discovered emission line features in the remnant spectrum. 
The feature at $E=4.0\pm 0.2$ keV is detected only in the northern 
rim. The width of this feature ($\sigma=0.3$~keV) is a little 
larger than the energy resolution of ACIS-I at 4~keV (i.e. $\sigma\sim0.1$~keV). 
This might indicate that the feature can be a blend of several lines.
Its energy centroid is close to a number of K emission lines 
from $^{44}$Ca and/or $^{44}$Sc (cf. Table 5 in Iyudin et al.~2005 and 
references therein). Therefore, we speculate that the feature at 4~keV 
can possibly comprises these lines. 

If the identification is correct, it suggests a possible 
presence of $^{44}$Ti because both $^{44}$Ca and $^{44}$Sc are produced in the 
decay chain $^{44}$Ti$\rightarrow$$^{44}$Sc$\rightarrow$$^{44}$Ca. 
The half-life of $^{44}$Ti is $\sim60$ yrs. If the line feature at 4 keV is
indeed from the decay of $^{44}$Ti, its short half-life implies that the remnant should be 
rather young. Also, the production of $^{44}$Ti in the supernova 
is very sensitive to the explosion mechanism and the ejecta dynamics (see 
discussion in Iyudin et al.~2005). Therefore, obtaining the yield and spatial 
distribution of $^{44}$Ti can help to further constrain the physical properties 
of the remnant as well as the nature of the progenitor. 

On the other hand, the broadening of the feature can also result from the motion of the 
ejecta. However, this would imply the ejecta to have a velocity of $\sim10000$~km/s. 
This is not consistent with \G\ being the result of a low energy supernova explosion 
as inferred from our X-ray data analysis as well from  optical observation. 
Furthermore, the limited energy resolution of imaging data precludes 
any accurate measurement of the line width. High resolution grating spectroscopy is 
required to disentangle the possible blend of lines as well as determine their widths. 

Apart from the feature at 4 keV, we have also found an emission line feature 
at $\sim7$ keV in both, the northern and southern rims. The large errors of 
the energy centroid and the width do not allow a firm identification. A deep 
observation is required to better constrain these parameters. 
We note that the approximate energy of the feature is rather close to the Fe-K emission. 
This leads us to speculate that the feature can be a result from unresolved 
Fe K$_\alpha$ and K$_\beta$ lines. This also requires data with high spectroscopic 
resolution for further investigation.  

The Chandra observation of \G\ also leads to the detection of X-ray point sources 
at the proximity of the remnant's geometrical center. However, the 
limited photon statistics do not allow a constraining spectral analysis of these 
sources. The non-detection of optical counterparts for \sten\ and \hard\
leaves these two objects as possible candiates for a stellar compact
remnant.  A dedicated deep optical observation is desirable to better 
constraining the nature of the point sources. 

\begin{acknowledgements}
CYH is supported by the Croucher Foundation Postdoctoral Fellowship. 
CYH would like to thank Herbert H.B.~Lau and J.J.~Eldridge for 
discussing the possible progenitors of low energy supernova. The authors 
would also like to thank the referee for providing many useful suggestions to
improve the quality of the manuscript. 
\end{acknowledgements}

\end{document}